\documentclass[3p,twocolumn,final]{elsarticle}
\makeatletter
\def\ps@pprintTitle{%
   \let\@oddhead\@empty
   \let\@evenhead\@empty
   \let\@oddfoot\@empty
   \let\@evenfoot\@oddfoot
}
\makeatother

\usepackage{lineno,hyperref}
\modulolinenumbers[5]

\usepackage{templates/numcompress}\biboptions{super}

\usepackage[printonlyused]{acronym}
\usepackage{framed}
\usepackage{soul}
\usepackage{graphicx,xcolor}
\usepackage{amsfonts}
\usepackage{amsmath,amssymb,gensymb}
\usepackage[mathcal]{euscript}
\usepackage{epstopdf}
\usepackage{enumitem}
\usepackage[labelsep=period,figurename=Fig.]{caption}
\usepackage{booktabs}
\usepackage{threeparttable}
\usepackage{adjustbox}
\usepackage{multirow}
\usepackage{nopageno}
\usepackage{fullpage}

\usepackage[version=3]{mhchem}
\usepackage{comment,cleveref}
\crefname{figure}{fig.}{figs.}
\Crefname{figure}{Fig.}{Figs.}

\acrodef{bdb}[BDB]{Beyond Design Basis}
\acrodef{cap}[CAP]{Condition Adverse to Quality Program}
\acrodef{cba}[CBA]{Cost-Benefit Analysis}
\acrodefplural{cba}[CBAs]{Cost-Benefit Analyses}
\acrodef{cdf}[CDF]{Core Damage Frequency}
\acrodef{chrs}[CHRS]{Containment Heat Removal System}
\acrodef{crmp}[CRMP]{Comprehensive Risk Management Process}
\acrodefplural{crmp}[CRMPs]{Comprehensive Risk Management Processes}
\acrodef{db}[DB]{Design-basis}
\acrodef{dba}[DBA]{Design Basis Accident}
\acrodef{did}[DID]{Defense-in-Depth}
\acrodef{eccs}[ECCS]{Emergency Core Cooling System}
\acrodef{esd}[ESD]{Event Sequence Diagram}
\acrodef{flc}[FC]{Fails Closed}
\acrodef{fda}[FDA]{Food and Drug Administration}
\acrodef{fmea}[FMEA]{Failure Modes and Effects Analysis}
\acrodefplural{fmea}[FMEA]{Failure Modes and Effects Analyses}
\acrodef{fsar}[FSAR]{Final Safety Analysis Report}
\acrodef{fs}[FoS]{Factor of Safety}
\acrodefplural{fs}[FsoS]{Factors of Safety}
\acrodef{ft}[FT]{Fault Tree}
\acrodef{fts}[FTS]{Fail To Start}
\acrodef{ftr}[FTR]{Fail To Run}
\acrodef{eno}[ENO]{Extraordinary Nuclear Occurrence}
\acrodef{pla}[PLA]{Public Liability Action}
\acrodef{gra}[GRA]{Generation Risk Assessment}
\acrodef{lerf}[LERF]{Large Early Release Frequency}
\acrodef{mss}[MSS]{Main Steam System}
\acrodef{mslb}[MSLB]{Main Steam Line Break}
\acrodef{nei}[NEI]{Nuclear Energy Institute} 
\acrodef{npv}[NPV]{Net Present Value}
\acrodef{npp}[NPP]{Nuclear Power Plant}
\acrodef{nrc}[NRC]{Nuclear Regulatory Commission}
\acrodef{oqap}[OQAP]{Operations Quality Assurance Program}
\acrodef{lar}[LAR]{License Amendment Requests}
\acrodef{lb}[LB]{Licensing Basis}
\acrodef{lerf}[LERF]{Large Early Release Frequency}
\acrodef{loca}[LOCA]{Loss of Coolant Accident}
\acrodef{lwr}[LWR]{Light Water Reactor}
\acrodef{om}[O\&M]{Operations and Maintenance}
\acrodef{ora}[ORA]{Organizational Risk Assessment}
\acrodef{pga}[PGA]{Peak Ground Acceleration}
\acrodef{pra}[PRA]{Probabilistic Risk Assessment}
\acrodef{pwr}[PWR]{Pressurized Water Reactor}
\acrodef{rcb}[RCB]{Reactor Containment Building}
\acrodef{rcd}[RCD]{Reactor Core Damage}
\acrodef{rcs}[RCS]{Reactor Coolant System}
\acrodef{rr}[RR]{Radiation Release}
\acrodef{ssc}[SSC]{Systems, Structures, and Components}
\acrodef{stm}[STM]{State Transition Matrix}
\acrodefplural{stm}[STMs]{State Transition Matrices}
\acrodef{ufsar}[UFSAR]{Updated Final Safety Analysis Report}
\acrodef{nrc}[NRC]{Nuclear Regulatory Commission}
\acrodef{doe}[DOE]{Department of Energy}
\acrodef{omb}[OMB]{Office of Management \& Budget}
\acrodef{oira}[OIRA]{Office of Information \& Regulatory Affairs}

\DeclareMathOperator*{\argmax}{arg\,max}
\DeclareMathOperator*{\argmin}{arg\,min}

\usepackage{sectsty}

\renewcommand{\thefootnote}{\Alph{footnote}}

\makeatletter
\renewcommand{\MaketitleBox}{%
  \resetTitleCounters
  \def\baselinestretch{1}%
  \begin{center}
    \def\baselinestretch{1}%
    \Large \@title \par
    \vskip 18pt
    \normalsize\elsauthors \par
    \vskip 10pt
    \footnotesize \itshape \elsaddress \par
  \end{center}
  \vskip 12pt
}
\makeatother

\begin{document}


\begin{frontmatter}

\title{Safety--Critical Protective Systems and Margins of Safety} 

\author{Martin Wortman\corref{correspondingauthor}}
\author{Ernie Kee}
\author{\& Pranav Kannan}
\address{The Organization for Public Awareness of Hazardous Technology Risks}






\end{frontmatter}

\renewcommand{\thefootnote}{\alph{footnote}}
\sectionfont{\normalfont\normalsize\bfseries}
\subsectionfont{\normalfont\normalsize\bfseries}

{\itshape

The design and operation of protective systems is an essential engineering responsibility.  Ensuring public safety, while essential, must be accomplished at a feasible cost and within government regulation. 
Hence, protective system design and operational decisions must be evaluated with respect to benefit (both enterprise profit and social benefit) and cost (both enterprise and social costs). 

Analytical arguments are made that establish the economic relationship between protective system margins of safety, regulatory authority, and the calculus of negligence. 
Within this risk-based analytical framework, protection efficacy is explored.
In particular, the risk-economics of margins of safety are examined by identifying the reference efficacy with respect to which margins of safety are measured. 
Engineering design and operations decisions intended to improve protection efficacy can, thus, be gauged as the degree to which they advance a risk-based margin of safety.

Finally, our analytical framework is exercised to show how risk-based margins of safety reveal the relationship between uncertain costs and regulatory activity focused on ensuring public welfare that is backstopped by liability in the event of catastrophe.  
How both prescriptive and performance based regulations influence  
margins of safety with respect to protective system innovation can be identified here.
}

\vspace{0.05in}


\section{TENETS of PROTECTIVE SYSTEM DESIGN}
	Protective measures are deployed almost without exception whenever technology operates in situations 
	where its failure can cause significant financial and/or physical harm. 
	Today, engineers are responsible for developing and operating sophisticated protective systems that 
	integrate hardware, personnel, and information to ensure public safety. 
	When protective systems fail to mitigate catastrophic events, liability must be determined through regulatory 
	response, judicial inquiry, or both. 
	It is every engineer's nightmare that she could be the source responsible for an error in design or operations 
	that causes a catastrophic protective system failure.
	
	\subsection{Profit, Design, and Hazard}
	Of particular interest are the design or operational errors made in the commercial nuclear power sector.
	Of course, commercial nuclear power utility engineers are responsible for developing economical 
	\emph{and} safe designs; a difficult task in the nuclear power sector.
	At this time, interest is focused on cost of production in the setting where there exists a social
	benefit (social welfare) and a level of hazard (such as catastrophic failure in commercial nuclear power);
	an important topic still lacking an academic basis has not been established.
	Ther focus is narrow, on the commercial nuclear power setting where regulation is enforced on many aspects of
	design and operation.
	Renewal following catastrophic failure is not attempted.
	This topic is important because the utility engineer is faced with balancing cost against a level of safety (hazard rate) 
	and, without better support, must proceed on a qualitative basis.
	We believe we can help her do better by moving to a more quantitative basis for support.
	That is, the framework in which the engineer must make design decisions normally lacks
	consideration of the ultimate goal to be achieved in the regulated setting.
	Absent a basic understanding of the elements required to quantify costs involved with design decisions,
	suboptimal designs are a likely outcome.
	
\begin{figure}[ht]
		\centering
		\includegraphics[width=0.45\textwidth]{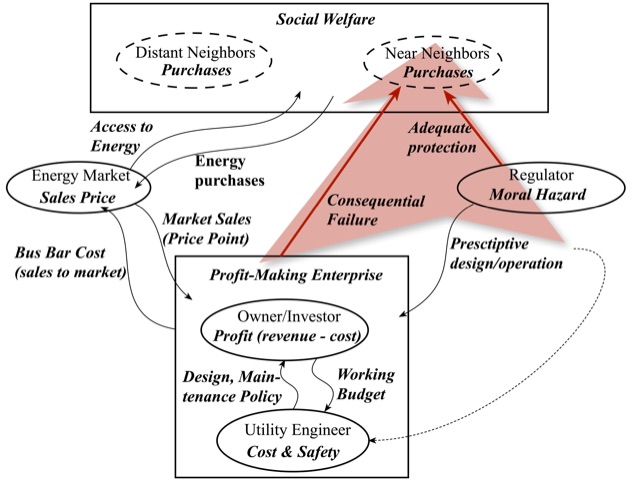}
		\caption{A graphical view of the relationships among the main elements, including the utility engineer, that need to be 
		considered in developing an understanding of the social welfare function for energy production
		by a profit-making enterprise.}
		\label{fig:socialgood}
	\end{figure}

	Quantitive support is especially important to for--profit enterprise where the engineer's 
	decisions are made in a very complex setting and many options are available.
	Some of the elements and interactions in \Cref{fig:socialgood} that come to mind are illustrated.
	When market conditions are unfavorable, the utility engineer experiences the tensions that arise in
	balancing the economic responsibility to the owners and investors (reducing operational and 
	maintenance costs) and the need for safe design and operation (presumably coming at higher cost).
		
	The regulator also must make decisions on ``adequate protection'', perhaps less directly than an engineer
	engaged in design and operation decision--making, which is, to some extent, based on governmental and public 
	feedback.
	For example, recent legislation\footnote{``Nuclear Energy Innovation and Modernization Act (S 2795)'' US Senate.} 
	ask for more expedient resolution by the \ac{nrc} on the issue of license actions.
	Although for now we leave quantitative study of the tension created by moral, economic, social, and regulatory 
	impacts on engineering design decision--making to future work, these relationships largely motivate what follows.
	

	
	\subsection{Summary}
	When thinking about `safety', the engineer generally has in mind a design point at which catastrophic failure 
	is less likely to occur than another design point (`unsafe design point'); designs less likely to fail 
	\emph{when compared to another design} are more `safe' states \citet{hansson2012}
	For example, in the use of a ``factor of safety.'' 
	We are particularly interested in designs that have more (or less) failure likelihood when compared to those 
	meeting regulatory standards.
	Recognition of such standards may be difficult to define 
	\citeauthor{Bjelland2013}\citep[][]{Bjelland2013} argues even though may make the argument 
	\ac{pra} makes these kinds of estimates.
	The regulatory ``adequate protection'', together with the design and operational decisions made by 
	the for--profit enterprise (largely an engineering responsibility) that can lead to catastrophic failure 
	(\Cref{fig:socialgood}) should minimize the harm to the public while maximizing the social welfare.
	Groundworkis being laid towards quantifying the economics involved in the decision--making process.
	
\section{REGULATORY AUTHORITY and the CALCULUS of NEGLIGENCE}
	
	\citet{000323324500002n.d.} 
	develops quantitative liability measures in a general setting.
	The examination examination of safety margin should be prefaced with observations about regulatory authority, 
	moral hazard, liability and the calculus of negligence.
	Moral hazard occurs when one party decides how much risk to take and another party pays if things go badly. 
	While tort law provides legal remedies for those who have been harmed through the negligence of another 
	party, it does not ensure prior protections from that harm.  
	The relationship between taking risks and negligence is codified in the so called Calculus of Negligence 
	(sometimes called the Hand Rule\footnote{The calculus of negligence, or ``Hand Rule", is a 
	legal standard applied to establish liability for harm. If the harm could be avoided for less than 
	the cost of the harm.}). 
	The Hand Rule provides an analytical definition of negligence and is often written as 

	$$b<P \dot L,$$

	\noindent where $\dot L$ is value of loss (harm) caused, $P$ is the probability of that loss, and  $b$ is the 
	burden of protection (value of the protection effort applied by the enterprise for the purpose of preventing harm). 
	The Hand Rule offers a litmus test to determine if adequate measures have been taken to prevent collateral harm. 
	However, liability for collateral harm does not eliminate moral hazard. 
	In fact, a profit maximizing enterprise that can potentially cause harm through negligence has no economic 
	incentive to provide protections beyond that which would marginally improve profit. 

	The general purpose of the government's regulatory authority is to mitigate moral hazard. 
	The development, deployment, and operation of protective systems overlaying nuclear facilities is, in part, 
	dictated by regulation. 
	Regulations serve to provide prior protections for the interests of neighbors who are not profit 
	claimants and who might be harmed should a given nuclear enterprise suffer a catastrophic event. 
	In this way, regulations do not provide for the general public benefit, but are directed at protecting those who 
	suffer moral hazard (for example neighbors).

	It is now observed that government regulations cause an enterprise to depart from the profit 
	maximization that might be achieved at an unregulated market equilibrium, and in doing cause that 
	enterprise to operate in a manner that is not economically risk--neutral.  
	Departure from risk--neutrality suggests that the risk preferences of the enterprise now reflect a risk--averse 
	social welfare responsibility.  
	In principle, when the enterprise preferences are well--ordered (rational) their regulated risk preferences 
	induce a social welfare function that is corollary to the Expected Utility Theorem.  
	The induced (enterprise--specific) social welfare function will be appealed to in the following arguments.
	
	\subsection{Negligence in the Commercial Nuclear Power Setting}
		The regulatory 
		framework in the nuclear industry is unique in the sense that the federal administrative agency has an 
		absolute say in how liability should be assigned in a nuclear incident. 
		Nuclear technology was largely a monopoly within the hands of the federal government until passage 
		of the Atomic Energy Act of 1946 (McMahon Act) where the federal government allowed private production 
		of nuclear energy via licensing and regulation.  
		Nevertheless, despite a regulatory incentive for investment, the private sector was reluctant to be involved in the 
		development of nuclear energy, largely due to the fear of incurring strict or unlimited liability in case of a nuclear incident, regardless 
		of how insignificant the incident.\footnote{Production of nuclear energy may be categorized as ``abnormally dangerous activities'' 
		and is thus subject to strict liability.
		Restatement of Torts Second, section 519, provides: (1) One who carries on an abnormally dangerous activity is subject to liability 
		for harm to the person, land or chattels of another resulting from the activity, although he has exercised the utmost care to prevent 
		the harm. (2) This strict liability is limited to the kind of harm, the possibility of which makes the activity abnormally dangerous.
		To determine whether an activity is abnormally dangerous, see Restatement of Torts Second, section 520.
		See also WILLIAM L. PROSSER, THE LAW OF TORTS, \S 78 at 516 (4th Ed. 1971) (``In the field of strict liability, the first case raising 
		the question as to the use of nuclear energy has yet to reach the courts. When it does, it may be predicted with a good deal of 
		confidence that this is an area in which no court will, at last, refuse to recognize and apply the principle of strict liabilityÉ'')}  
		
		To address industry concerns and facilitate development of the then--fledgling commercial nuclear power industry, the Price--Anderson Act 
		was enacted into law in 1957.
		This sets a limit on the monetary liability of private producers of nuclear power in case of a catastrophic 
		nuclear accident, and defines the procedural mechanisms for insurance coverage in the nuclear industry.\footnote{\emph{See} Dubin, 
		Jeffrey A., and Geoffrey S. Rothwell. ``SUBSIDY TO NUCLEAR POWER THROUGH PRICE-ANDERSON LIABILITY LIMIT.'' Contemporary 
		Economic Policy 8.3 (1990): 73--79. Between 1959 and 1982, the Act set a limit of \$560 million on the liability of civilian operators of nuclear 
		power plants for accidental damages. After the 1988 amendment, this limit was increased to \$7 billion.}  
		In the 1988 amendment to the Price--Anderson Act, the Congress established a sole and exclusive federal cause of action known as a 
		\ac{pla}, for ``any legal liability arising out of or resulting from a nuclear incident or precautionary evaluation''\footnote{In 
		re TMI Litigation Cases Consol. II, 940 F.2d 832 (3d Cir. 1991).}, 
		thereby preempting all state cause of actions for damages arising from nuclear incidents covered under the Act.\footnote{See In re TMI Litig. 
		Cases Consol. II, 940 F.2d 832, 855 (3d Cir. 1991) (``In explicitly providing that the `substantive rules for decision' in public liability actions `shall 
		be derived from' the law of the state in which the nuclear incident occurred, . . . Congress expressed its intention that state law provides the 
		content of and operates as federal law.''); O'Conner v. Commonwealth Edison Co., 13 F.3d 1090, 1099-1100 (7th Cir. 1994) (``Thus, a state 
		cause of action is not merely transferred to federal court; instead, a new federal cause of action supplants the prior state finding.}
	
		Whether or not implementation of protective systems can save private operators from being held liable for personal injury and property 
		damage arising from the nuclear incidents might depend on the actual extent of the incident, that is, whether the event is an \ac{eno} or not. 
		An \ac{eno} is defined as ``any event causing a discharge or dispersal of source, special nuclear, or byproduct material from its intended 
		place of confinement in amounts offsite, or causing radiation levels offsite, which the \ac{nrc} or the Secretary of 
		Energy, as appropriate, determines to be substantial, and which the \ac{nrc} or the Secretary of Energy, as appropriate, determines has 
		resulted or will probably result in substantial damages to persons offsite or property offsite.''\footnote{42 U.S.C. \S 2014.}  
		This determination made by the \ac{nrc} of the Secretary of Energy shall be final and conclusive, and is not reviewable by 
		the judicial system.\footnote{42 U.S.C. \S 2014(j).}
		
		The elements for establishing a \ac{pla} cause of action for an \ac{eno} case are essentially the same as the ones for negligence under state 
		law, that is, the presence of a duty of care, breach of duty, proximate cause and damages. 
		Nevertheless, if a nuclear incident is determined by \ac{nrc} or \ac{doe} to qualify as an \ac{eno}, it can be almost certain that the \ac{pla} 
		duty has already been breached (that is, the radiation released has exceeded the federal dose limits), which means the first two 
		elements of \ac{pla} have been established to be adverse to the defendant (the operator). 
		To speed up the litigation process, the \ac{nrc} or \ac{doe} then may, but not necessarily, require the defendant in the \ac{pla} lawsuit to 
		waive the defense of not being at fault (breach of duty owned),\footnote{42 U.S.C. \S 2210(n). } 
		creating a circumstance that is similar to the common law strict liability, except that such waiver is controlled by the federal agency 
		and cannot be overruled by the court. 
		By doing so, the federal government can effectively prevent spending scarce judicial resources on litigation over the elements 
		of cause of action that are not questionable and squandering the limited financial protection provided by Price--Anderson.\footnote{Under 
		Price--Anderson Act, all defense costs are subtracted from the available funds (``financial protection'') for compensation.} 
		Accordingly, in case of an \ac{eno}, whether a protective system has been installed or not can hardly alter the facts that the \ac{pla} duty 
		has been breached, especially when a waiver of defense is forced upon the defendant. However, if the protective measures have 
		reduced the damages resulting from the nuclear incident in practice, it could still be instrumental to the operator since it can lower 
		the amount of compensation the insurance company/operator needs to pay.
		
		\paragraph{Summary}
		How an \ac{eno} case would proceed in court remains a scenario in theory since an \ac{eno} has never occurred. 
		According to the \ac{nrc}, even the Three Mile Island accident did not release enough radiation to qualify as an \ac{eno}. 
		Therefore, in reality, almost all nuclear accidents would fall under the category of non--\acp{eno}. 
		The \ac{pla} for non--\acp{eno} have the same four elements as \ac{eno} cases. 
		But here, adoption of protective measures may be decisive if they keep release of radioactive materials (exposures) under the federal 
		radiation safety standards. 
		In such cases, the defendant might be exempted from legal liability as no duty has been breached.  
		Generally speaking, a non--\ac{eno} case may be litigated in court as a typical negligence in court.

	\subsection{Information and Design}
		The uncertainties associated with the efficacy of protection are difficult to identity and (when identified) 
		extremely difficult to quantify. 
		Protective systems are exceptional in the endeavor of engineering design in that they must be created and 
		operated, while ALL stakeholders hope that the circumstances for which the protection is deployed will never 
		actually occur. 
		It is well-understood that the most reliable conceivable safety technologies and protective system are 
		rarely deployed because of cost; there always exist tradeoffs between safety and expense; hence, 
		protective systems are necessarily the focus of much mathematical and empirical engineering analysis. 
		Stress--testing protective systems with real--life catastrophes is obviously not a desirable source of 
		information in engineering design (although some testing on subsystems may be done). 
		Further, engineers cannot avoid design decisions simply because operational experience 
		(for example catastrophic event data) is limited. 
		In the absence of rich operational data, it is important to address uncertainty in a manner consistent 
		with the best available theory and analytics. 
		When the stakes are truly high, there is no room for \emph{ad hoc} methods or anecdotal reasoning.

		In the developments that follow, the authors appeal to the well--accepted tenets of utility theory and the 
		Kolmorogov axiomization of probability measure.
		This places the authors in the mainstream of established economic theory, mathematics, and epistemology. 
		Arguments will be built beginning with the expected utility theorem and results developed that 
		reveal the relationship between margin of safety, information, and prudent engineering decisions in 
		the design and operation of protective systems.

		Consider a situation where a regulated enterprise will deploy a technology that has been selected 
		on the basis of public safety requirements, revenue, and social costs; generally regulatory constraints 
		are enforced that cause the enterprise to find the enhanced safety of the selection preferable to all 
		other candidate technologies that could serve (the identical) future demand trajectories. 
		In what follows, the ``\emph{margin of safety}" associated with the preferred selection will be explored and the
		technology selection decision will be assumed to follow all tenets of the 
		\emph{von Neumann -- Morgenstern Expected Utility Theorem}. 
		All candidate technologies would face identical future demand, generating identical revenue trajectories 
		so long as the technology does not suffer catastrophic failure, ending its useful life. 
		While the lifecycle costs can differ significantly among various technology alternatives, lifecycle 
		revenues are assumed to be identical. 
		The regulatory environment is captured through social welfare function $u: \mathbb{R} 
		\rightarrow \mathbb{R}$ on the value of all alternative technologies.
		\bigskip

	\subsection{Analytical Framework}
		In the arguments to follow, we have need to identify multiple probability spaces. 
		In the interest of manageable notation, the de Finetti notation is adopted. 
		Here, for a random variable $X$ defined on the probability space $(\Omega, {\cal{F}}, P),$ the 
		traditional expectation integral $E[X]$ is replaced by $P(X)$.

		Let all candidate technologies, available for possible selection by the enterprise, be indexed with indices 
		belonging to the set ${\cal{A}}$ where $\alpha^* \in {\cal{A}}$ is the preferred technology. 
		Thus, there is a collection of probability spaces $\{(\Omega_\alpha, {\cal{F}}_\alpha, P_\alpha);  
		\alpha \in {\cal{A}}\}.$  For each alternative $\alpha \in \ {\cal{A}},$ define on $\{(\Omega_\alpha, 
		{\cal{F}}_\alpha, P_\alpha)$ the random variables:
 
 		$V_\alpha: \Omega_\alpha \rightarrow  \mathbb{R}$, the net present value of technology 
		alternative $\alpha$,
 
 		$C_\alpha: \Omega_\alpha \rightarrow  \mathbb{R}_+$, the lifecycle cost of alternative $\alpha$,
 
 		$\chi_\alpha: \Omega_\alpha \rightarrow  \{0, 1\}$, where $\chi_\alpha = 1$ on the event that the 
		lifetime of alternative $\alpha$ terminates in catastrophe.
 
 		\noindent
 		Inasmuch as the enterprise has rationally selected technology alternative $\alpha^* \in {\cal{A}}$, 
		it follows from the expected utility theorem that 
	
		 \begin{equation}\label{EUT}
 			\alpha^* = \argmax_{\alpha \in {\cal{A}}} E_\alpha[u \circ V_\alpha].
 		\end{equation}
		Note that, since 
		any selected technology must follow the same demand trajectory, $V_\alpha = -C_\alpha$,  
		$\forall \alpha \in {\cal{A}}.$  Hence, it follows that \eqref{EUT} can be rewritten as 
		$$\alpha^* = \argmin_{\alpha \in {\cal{A}}} E_\alpha[u \circ C_\alpha]$$
		where, $E_\alpha[u \circ V_\alpha]$ is the \emph{expected lifecycle social cost} of technology 
		alternative $\alpha \in {\cal{A}}$.

		It is important to recall that technology $\alpha^*$ is selected because regulation has imposed a 
		value on public safety (implicitly represented by the social welfare mapping $u$), which reflects the 
		high social cost associated with catastrophic failures that terminate a technology's lifecycle. 
		Thus, is it useful to explore lifecycle social costs on catastrophic events. 
		In this way, the margin of safety that certain non-optimal alternatives might enjoy over  $\alpha^*$ can be investigated. 
		To this end, note that the expected lifecycle social cost can be written as, 
		$$E_\alpha[u \circ C_\alpha] = E_\alpha[E_\alpha[u \circ C_\alpha | \chi_\alpha]], \forall \alpha \in {\cal{A}},$$
 		or
 		\begin{equation} \label{nested}
		\begin{split}
 			& E_\alpha[u \circ C_\alpha] = E_\alpha[u \circ C_\alpha | \chi_\alpha = 0]P_\alpha(\chi_\alpha = 0) + \\
			& E_\alpha(u \circ C_\alpha | \chi_\alpha = 1)P_\alpha(\chi_\alpha = 1).
			\end{split}
		\end{equation}

		\noindent
		As a matter of convenience, the following is defined: 

		$c_\alpha^g \triangleq E_\alpha(u \circ C_\alpha | \chi_\alpha = 0)$, the expected social cost of 
		alternative $\alpha$ on the event that lifecycle terminates \emph{without catastrophe},

		$c_\alpha^f \triangleq E_\alpha(u \circ C_\alpha | \chi_\alpha = 1)$, the expected social cost of 
		catastrophe--free lifecycle, 

		\noindent
		and,

		$p_\alpha \triangleq P_\alpha(\chi_\alpha = 1),$ $\alpha \in {\cal{A}}.$

		\noindent
		Hence, \eqref{nested} is rewritten as
 		\begin{equation} \label{fundamental}
 			E_\alpha(u \circ C_\alpha) = c_\alpha^g  + (c_\alpha^f - c_\alpha^g)p_\alpha ,  \forall \alpha \in {\cal{A}}.
 		\end{equation}

		\noindent
		$c_\alpha^p \triangleq (c_\alpha^f - c_\alpha^g)$ will be referred to as as the \emph{catastrophe--premium} 
		of technology $\alpha$. 
		Thus, \eqref{fundamental} states that: 

		\noindent
		\emph{\textbf{For any technology alternative, its expected social cost is given by its expected social cost 
		with catastrophe--free operation, plus its catastrophe--premium weighted by the probability of catastrophe.}}

		It now follows from \eqref{EUT} and \eqref{fundamental} that for all $\alpha \ne {\alpha^*}$
		\begin{equation}
  			c_{\alpha^*}^g  + (c_{\alpha^*}^f - c_{\alpha^*}^g)p_{\alpha^*} \le c_\alpha^g + 
			(c_\alpha^f - c_\alpha^g)p_\alpha \nonumber
		\end{equation}
	
		or,
	
		\begin{equation} \label{ineq}
 			c_{\alpha^*}^g  + c_{\alpha^*}^p p_{\alpha^*} \le c_\alpha^g  + c_\alpha^p p_\alpha.
		\end{equation}
	
		\noindent Rearranging \eqref{ineq} into point--slope form gives
		\begin{equation*}
			p_{\alpha^*} \le \frac{c_\alpha^p}{c_{\alpha^*}^p} p_\alpha - \frac{(c_{\alpha^*}^g - 
			c_{\alpha}^g)}{c_{\alpha^*}^p}.\nonumber
		\end{equation*}
		$c_{({\alpha^*},\alpha)}^p \triangleq (c_{\alpha^*}^g - c_{\alpha}^g)$, the expected 
		difference in social cost  between technology alternatives $\alpha^*$ and $\alpha$, is referred in here as the 
		\emph{reliability premium} of  choosing $\alpha^*$ over $\alpha \in {\cal{A}}$. 
		Note that it may happen that the reliability premium takes a negative value (as would be the case 
		of rejecting a more reliable alternative because of its cost).
		Thus, it now follows that for all technology alternatives $\alpha \in {\cal{A}}$, 
		\begin{equation} \label{final}
			p_{\alpha^*} \le \frac{c_\alpha^p}{c_{\alpha^*}^p} p_\alpha - \frac{c_{(\alpha^*,\alpha)}^p}{c_{\alpha^*}^p}.
		\end{equation}

		\begin{figure}[t]
			\centering
			\includegraphics[width=0.47\textwidth]{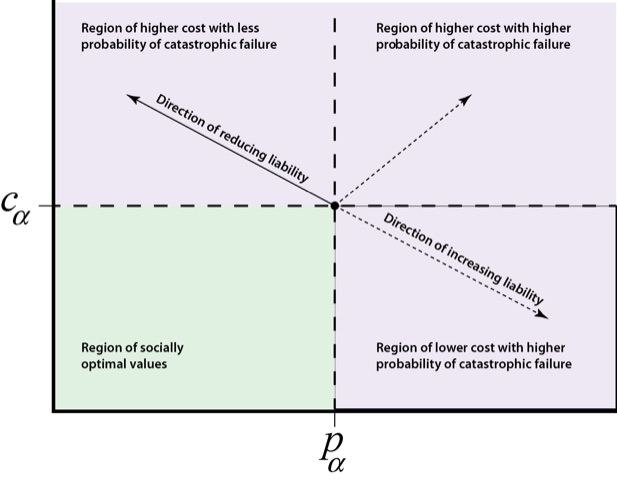}
			\caption{The nature of alternative selections in relation to cost ($C_{\alpha}$) and catastrophic failure probability ($p_{\alpha}$) 
			over the plant lifetime with relation to possible liability in light of failures.}
			\label{fig:point_slope}
		\end{figure}

		\begin{figure}[t]
			\centering
			\includegraphics[width=0.47\textwidth]{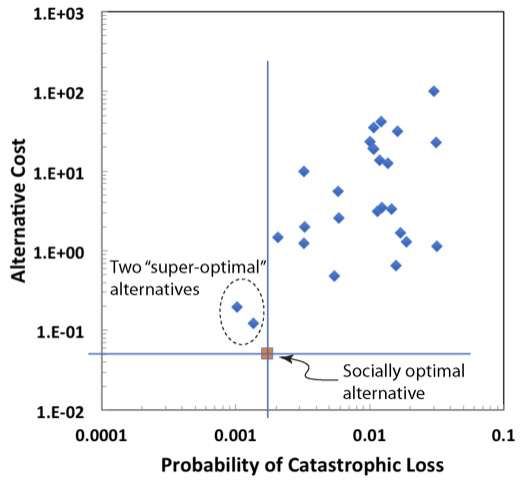}
			\caption{Twenty--six hypothetical socially suboptimal alternatives plotted with the socially optimal alternative selected from a total
			of twenty--seven alternatives hypotesized.}
			\label{fig:pcplot}
		\end{figure}

		To illustrate different aspects of \eqref{final} and complexity between socially and regulatory optimums, \Cref{fig:pcplot} is 
		created based on an ad hoc correlation between the probability of catastrophic failure and social costs for 27 hypothetical alternatives. 
		In the assumed correlation, the tendency is to relate smaller social costs with smaller probabilities of catastrophic failures which 
		would be desirable to industry and the regulator. 
		The figure shows that a case which is socially optimum would not necessarily be the regulatory optimum (``super--optimal''), since there 
		are two alternative technologies plotted to the ``northwest'' of it. 
		In this figure, super--optimal technologies would relate to alternatives selected under regulation (their corresponding 
		probabilities are less than the socially optimal one). 
		Once the socially optimal alternative is known, none of the alternatives would lie to the south of it.
		
		Engineers typically couch technology choices in terms of system reliability and cost. 
		\eqref{final} shows that $\alpha^*$ is the most preferred technology only when its life cycle unreliability 
		$p_{\alpha^*}$ is at least as small as the life cycle unreliability $p_{\alpha}$, for all $\alpha \in {\cal{A}},$ 
		scaled by the quotient of catastrophe premiums less the quotient of the reliability premium to the un--preferred 
		alternative's catastrophe premium.
		\Cref{fig:point_slope} illustrates the behaviors described by \eqref{final}.
		Thus, the conditions set forth by the expected utility theorem can be understood in terms that are both 
		analytically and intuitively specific to protective system design and operation. 
		Of course, in practice, the particular values of elements that form \eqref{final} are difficult to obtain 
		since information (including event probabilities and the social welfare function) is typically vague or incomplete. 
		Nonetheless, the design decision of selecting the most preferred technology alternative cannot be avoided.

\section{CONCLUSION}
	The examination of protective systems offered establishes a decision--analytical framework capturing  
	the relationship between margins of safety and regulatory authority.
	It is argued that because potential liability (as identified through the calculus of negligence and following from the well--known 
	Coase Theorem) does not substantially influence profit maximizing decisions associated with the design and operation of 
	safety--critical protective systems, regulatory authority necessarily arises so as to ensure mitigation of moral hazard for a 
	certain element of the public (those having large potential for losses in the event of a catastrophe). 
	Regulatory authority induces a unique (up to affine transformation  as corollary to the Expected Utility Theorem) social welfare 
	function that  enforces unique socially optimal price--point for regulated protection that does not enhance revenues. 
	Margins of safety are thus defined to be associated with protective system alternatives that exhibit a lower probability of 
	catastrophe than a unique socially--optimal level of protection. 
	The framework identifies reliability premiums and catastrophe premiums associated with safety margins in a manner that allows 
	protective system design and operation decisions to be considered in the context of system lifecycle expects costs.

\begin{center}REFERENCES\end{center}
\bibliography{BibTeX/bibtex_library_dec9.bib}
\clearpage
\end{document}